\documentclass[12pt]{spieman}  
\usepackage{amsmath,amsfonts,amssymb}
\usepackage{graphicx}
\usepackage{setspace}
\usepackage{tocloft}
\usepackage{MnSymbol}

\newcommand{\aap}{Astronomy and Astrophysics}
\newcommand{\pasj}{Publications of the Astronomical Society of Japan}

\title{Detailed Design of the Science Operations for the XRISM mission}
\author[a,b,*]{Yukikatsu Terada}
\author[c]{Matt Holland}
\author[c]{Michael Loewenstein}
\author[a,b]{Makoto Tashiro}
\author[d]{Hiromitsu Takahashi}
\author[e]{Masayoshi Nobukawa}
\author[d]{Tsunefumi Mizuno}
\author[b]{Takayuki Tamura}
\author[f]{Shin'ichiro Uno}
\author[b]{Shin Watanabe}
\author[c]{Chris Baluta}
\author[c]{Laura Burns}
\author[b]{Ken Ebisawa}
\author[g]{Satoshi Eguchi}
\author[d]{Yasushi Fukazawa}
\author[b]{Katsuhiro Hayashi}
\author[b]{Ryo Iizuka}
\author[a]{Satoru Katsuda}
\author[h]{Takao Kitaguchi}
\author[i]{Aya Kubota}
\author[j]{Eric Miller}
\author[c]{Koji Mukai}
\author[b]{Shinya Nakashima}
\author[k]{Kazuhiro Nakazawa}
\author[l]{Hirokazu Odaka}
\author[d]{Masanori Ohno}
\author[m]{Naomi Ota}
\author[b]{Rie Sato}
\author[h]{Makoto Sawada}
\author[b]{Yasuharu Sugawara}
\author[n]{Megumi Shidatsu}
\author[l]{Tsubasa Tamba}
\author[l]{Atsushi Tanimoto}
\author[n]{Yuichi Terashima}
\author[o]{Yohko Tsuboi}
\author[d]{Yuusuke Uchida}
\author[p]{Hideki Uchiyama}
\author[m]{Shigeo Yamauchi}
\author[c]{Tahir Yaqoob}
\affil[a]{Saitama University, Graduate School of Science and Engineering, Saitama-shi, Saitama, Japan}
\affil[b]{Japan Aerospace Exploration Agency, Institute of Space and Astronautical Science, Sagamihara, Kanagawa, Japan}
\affil[c]{National Aeronautics and Space Administration, Goddard Space Flight Center, Greenbelt, Maryland, United States}
\affil[d]{Hiroshima University, School of Science, Higashi-Hiroshima, Hiroshima, Japan}
\affil[e]{Nara University of Education, Department of Teacher Training and School Education, Nara-shi, Nara, Japan}
\affil[f]{Nihon Fukushi University, Faculty of Health Sciences, Handa, Aichi, Japan}
\affil[g]{Fukuoka University, Faculty of Science Department of Applied Physics, Fukuoka-shi, Fukuoka, Japan}
\affil[h]{RIKEN, Nishina Center, Wako, Saitama, Japan}
\affil[i]{Shibaura Institute of Technology, Department of Electronic Information Systems, Saitama-shi, Saitama, Japan}
\affil[j]{Massachusetts Institute of Technology, Kavli Institute for Astrophysics and Space Research, Cambridge, Massachusetts, United States}
\affil[k]{Nagoya University, Department of Physics, Nagoya, Aichi, Japan}  
\affil[l]{The University of Tokyo, Department of Physics, Bunkyo, Tokyo, Japan}
\affil[m]{Nara Women's University, Department of Physics, Nara-shi, Nara, Japan}
\affil[n]{Ehime University, Department of Physics, Matsuyama-shi, Ehime, Japan}
\affil[o]{Chuo University, Department of Physics, Bunkyo, Tokyo, Japan}
\affil[p]{Shizuoka University, Faculty of Education, Shizuoka-shi, Shizuoka, Japan}

\cftpagenumbersoff{figure}
\cftpagenumbersoff{table} 

\begin{document} 
\maketitle

\begin{abstract}
{\it XRISM} is an X-ray astronomical mission led by the Japan Aerospace Exploration Agency (JAXA) and National Aeronautics and Space Administration (NASA), with collaboration from the European Space Agency (ESA)
and other international participants, that is planned for launch in 2022 (Japanese fiscal year),
to quickly restore high-resolution X-ray spectroscopy of
astrophysical objects using the micro-calorimeter array
after the loss of {\it Hitomi} satellite.
In order to enhance the scientific outputs of the mission,
the Science Operations Team (SOT) is structured independently
from the instrument teams and the Mission Operations Team (MOT).
The responsibilities of the SOT are divided into four categories:
1) guest observer program and data distributions,
2) distribution of analysis software and the calibration database,
3) guest observer support activities,
and 4) performance verification and optimization activities.
Before constructing the operations concept of the {\it XRISM} mission,
lessons on the science operations learned from past Japanese X-ray missions 
({\it ASCA}, {\it Suzaku}, and {\it Hitomi}) are reviewed,
and 15 kinds of lessons are identified by categories,
such as lessons on the importance of avoiding nonpublic (``animal'') tools,
coding quality of public tools in terms of both the engineering viewpoint and calibration accuracy,
tight communications with instrument teams and operations teams,
well-defined task division between scientists and engineers, etc.
Among these lessons, 
a) the importance of early preparation of the operations from the ground stage, 
b) construction of an independent team for science operations separate from the instrument development, 
and c) operations with well-defined duties by appointed members 
are recognized as key lessons for {\it XRISM}.
Based on this, i) the task division between the mission and science operations and ii) the subgroup structure within the {\it XRISM} team are defined in detail as the {\it XRISM} Operations Concept.
Based on this operations concept, the detailed plan of the science operations is designed as follows.
The science operations tasks are shared among Japan, the US, and Europe and are performed by three centers: 
the Science Operations Center (SOC) at JAXA, the Science Data Center (SDC) at NASA, and European Space Astronomy Centre (ESAC) at the ESA.
The SOT is defined as a combination of the SOC and SDC. The SOC is designed to perform tasks close to the spacecraft operations, such as spacecraft planning of science targets, quick-look health checks, pre-pipeline data processing, etc., and the SDC covers tasks regarding data calibration processing (pipeline processing), maintenance of analysis tools, etc. 
The data-archive and user-support activities are planned to be covered both by the SOC and SDC.
Finally, the details of the science operations tasks and the tools for science operations are defined and prepared before launch.
This information is expected to be helpful for the construction of science operations of future X-ray missions.
\end{abstract}


{\noindent \footnotesize\textbf{*}Yukikatsu Terada,  \linkable{terada@mail.saitama-u.ac.jp} }

\begin{spacing}{2}   

\section{Introduction}
\label{sec:intro}
The X-Ray Imaging and Spectroscopy Mission ({\it XRISM})\cite{2018SPIE10699E..22T} is an X-ray astronomical mission led by the Japan Aerospace Exploration Agency (JAXA) and National Aeronautics and Space Administration (NASA), in collaboration with the European Space Agency (ESA) and other international partners, that is planned for launch in 2022 (Japanese fiscal year) to restore high-resolution X-ray spectroscopy after the loss of the {\it Hitomi} satellite\cite{2018JATIS...4b1402T}.
The {\it XRISM} mission has four scientific objectives\cite{2020SPIE_XRISM_Summary}: 1) understanding the formation of the structure of the universe and evolution of clusters of galaxies by measuring turbulent and Doppler velocities at the 300 km/s level in spatially resolved spectroscopy of clusters of galaxies, 2) understanding the circulation history of baryonic matter in the universe from high-resolution spectroscopy of phenomena such as supernova remnants and supernovae, 3) understanding the transport and circulation of energy in the universe by observing feedback from active galactic nuclei or outflow from super-massive black holes via high-resolution spectroscopy, and 4) new science based on unprecedented high-resolution X-ray spectroscopy, such as detailed diagnostics of collisional ionization and photo-ionized plasma. 
To meet these scientific objectives of the {\it XRISM} mission, the spacecraft and ground systems are designed to use the X-ray micro-calorimeter array Resolve and the X-ray CCD camera Xtend on the focal planes of the X-ray mirrors, which provide X-ray spectroscopy with a high-energy resolution of $\le$ 7 eV FWHM within a field of view (FOV) of $2.9 \times 2.9$ arcmin$^2$ and imaging capability with a wide FOV of $30 \times 30$ arcmin$^2$, respectively, in the 0.3 to 12 keV band.
This paper focuses on the ground systems of the {\it XRISM} mission.

In order to maximize the scientific outputs from the {\it XRISM} mission, the science operations of the mission also need to be well designed and performed properly, namely, by conducting a guest observation program operating as a public observatory under a well-supported system of guest observers and providing well-calibrated observational data in the standard format for astronomical use (i.e., flexible image transport system [FITS] format \cite{2001A&A...376..359H}) with simple and accurate analysis environments and tools \cite{2018JATIS...4a1207A}. 

This paper aims to describe the details of the development of the {\it XRISM} Science Operations from the concept study to the detailed plans, as well as give detailed descriptions of the preparations for the operation (such as science operations manuals, tools, websites, etc) based on the SPIE Proceeding in 2020\cite{2020SPIE_XRISM_SOT}.
Note that such descriptions on the detail of design of the science operations may have sensitive topics for the project but the paper aims to describe those as much as possible avoiding confidential technical ideas and political issues for the {\it XRISM} project and agencies, because the authors believe that this knowledge may help the design of science operations in near-future high-energy missions. 
The rest of this paper is organized as follows.
We summarize the lessons learned from past X-ray missions in Section \ref{sec:ll} as the first step of the concept study, and summarize the concept of the operations in Section \ref{sec:concept}.
In Sections \ref{sec:team} and \ref{sec:plan}, the team structure and the details of the science operations plan are summarized, respectively.
Finally, Section \ref{sec:prep} describes the timeline of the science operations and details of preparation of tools in the ground systems for the {\it XRISM} Science Operations, and finally we summarize this paper in Section \ref{sec:summary}.

\section{Lessons for Science Operations Learned from Previous X-ray Missions}
\label{sec:ll}

\subsection{Summary of Lessons and Their Relations}
\label{sec:ll:summary}
As described in Section \ref{sec:intro}, the goal of science operations is to enhance or maximize the scientific outputs of the mission.
In science missions, the activities required of science operations can be divided into the following four categories.
\begin{quote}
    \begin{description}
    \item[SO1] Guest Observer (GO) program and data distribution
    \item[SO2] Distribution of analysis software and calibration database 
    \item[SO3] GO supporting activities
    \item[SO4] Performance verification and optimization (PVO) activities
    \end{description}
\end{quote}

Many lessons were learned from the science operations in the series of Japanese X-ray satellites, and although some of them require no changes, others need to be addressed before the next mission. 
Table \ref{tab:ll_summary} summarizes the relations among lessons learned from the Advanced Satellite for Cosmology and Astrophysics ({\it ASCA}\cite{1994PASJ...46L..37T}), {\it Suzaku}\cite{2007PASJ...59S...1M}, and {\it Hitomi}\cite{2018JATIS...4b1402T} missions, the details of which are described in Sections \ref{sec:ll:asca}, \ref{sec:ll:suzaku}, and \ref{sec:ll:hitomi} below.
Positive and negative lessons are marked by {\bf +} and {\bf -} identifiers, respectively.
Historically, attempts were made to address negative lessons in the next mission. However, this sometimes created another negative situation, which then needed to be solved in a subsequent mission. 
All of the lessons learned from past X-ray missions were considered by {\it XRISM} Science Operations, which are also shown in Table \ref{tab:ll_summary} and summarized in Section \ref{sec:ll:xrism} below.

\begin{table}[ht]
    \centering
    \caption{Relations among the lessons learned from {\it ASCA}, {\it Suzaku}, and {\it Hitomi}, summarized by category ({\bf SO1, SO2, SO3}, and {\bf SO4}; Section  \ref{sec:ll:summary}). Identifiers such as {\bf 1a-ASCA$^{+}$} and {\bf 2ab-Suzaku$^{+}$} are defined in Sections \ref{sec:ll:asca}, \ref{sec:ll:suzaku}, and \ref{sec:ll:hitomi}. The "$\rightarrow$" mark represents that the following mission continued the activities in the column on the left.}    
    \begin{tabular}{|c|c|c|c|c|}
    \hline 
    {\bf Category} & {\it\bf ASCA}      & {\it\bf Suzaku}         & {\it\bf Hitomi}       & {\it\bf XRISM}\\
    \hline 
    {\bf SO1} & {\bf 1a-ASCA$^{+}$}     & $\rightarrow$           &{\bf 1a-Hitomi$^\pm$}  & $\rightarrow$ \\ 
              &  (GO program)           &                         & (not activated)       &               \\ \cline{2-5}
              & {\bf 1b-ASCA$^{+}$}     & $\rightarrow$           & $\rightarrow$         & $\rightarrow$ \\
              & (data distribution)     &                         &                       &               \\ 
    \hline 
    {\bf SO2} & {\bf 2a-ASCA$^{+}$}     &  {\bf 2ab-Suzaku$^{+}$} & $\rightarrow$         & $\rightarrow$ \\ 
              & (tool verification)     &   (verification,        &                       &               \\ \cline{2-2} \cline{4-5}
              & {\bf 2b-ASCA$^{-}$}     &   public tools)         & $\rightarrow$         & $\rightarrow$ \\ 
              & (non-public tools)      &                         &                       &  \\ \cline{3-5}
              &                         & {\bf 2c-Suzaku$^{-}$}   & {\bf 2c-Hitomi$^{+}$} & $\rightarrow$ \\ 
              &                         & (software development)  & (specific team) & \\ \cline{4-5}
              &                         &                         & {\bf 2d-Hitomi$^{-}$} & to be fixed    \\ 
              &                         &                         & (management)     & \\
    \hline 
    {\bf SO3} & {\bf 3a-ASCA$^{+}$}     & $\rightarrow$           &                       & $\rightarrow$ \\ 
              &  (GO support)           &                         &                       &  \\ \cline{2-3} \cline{5-5}
              & {\bf 3b-ASCA$^{-}$}     & {\bf 3b-Suzaku$^{+}$}   & {\bf 3-Hitomi$^\pm$}  & $\rightarrow$ \\ 
              &  (support center)    &  (Help desk Japan)      &  (not activated)      &               \\ \cline{3-3} \cline{5-5}
              &                         & {\bf 3c-Suzaku$^{-}$}   &                       & to be fixed    \\ 
              &                         &  (data access rights)   &                       &               \\ \cline{5-5} \cline{5-5}
    \hline 
    {\bf SO4} & {\bf 4a-ASCA$^{+}$}     & $\rightarrow$           & $\rightarrow$         & $\rightarrow$ \\ 
              & (communication          &                         & (communication        & \\ 
              &  Japan/US)              &                         &  Japan/US/ESA)        & \\ \cline{2-5}
              & {\bf 4b-ASCA$^{+}$}     & $\rightarrow$           & {\bf 4b-Hitomi$^\pm$} & $\rightarrow$  \\ 
              & (calibration)           &                         &  (limited calibration)& \\ \cline{3-5} 
              &                         & {\bf 4c-Suzaku$^{+}$}   & $\rightarrow$         & $\rightarrow$ \\ 
              &                         & (IACHEC)                &                       &\\
    \hline 
    \end{tabular}
    \label{tab:ll_summary}
\end{table}

\subsection{Lessons Learned from {\it ASCA}}
\label{sec:ll:asca}
The {\it ASCA} mission was the fourth in the series of Japanese X-ray satellites \cite{1994PASJ...46L..37T} and was launched in 1993 carrying a Gas Imaging Spectrometer and Solid-state Imaging Spectrometer X-ray CCD cameras to observe astrophysical objects in the 0.5--10 keV band.
The science operations activities as a public observatory were well established in almost the first collaboration between the Institute of Space and Astronautical Science (ISAS) at current JAXA and NASA/GSFC in the GO program ({\bf 1a-ASCA$^{+}$}) and the distribution of observation data ({\bf 1b-ASCA$^{+}$}), analysis software ({\bf 2a-ASCA$^{+}$}), GO support ({\bf 3a-ASCA$^{+}$}), international collaboration ({\bf 4a-ASCA$^{+}$}), and calibration of instruments ({\bf 4b-ASCA$^{+}$}), but there were also two negative items ({\bf 2b-ASCA$^{-}$} and {\bf 3b-ASCA$^{-}$}).
The successful parts of {\it ASCA} are summarized below.
\begin{quote}
    \begin{description}
    \item[1a-ASCA$^{+}$:] The GO program worked well both in Japan and the US. GOs were able to submit their proposals to agencies, which were reviewed by the scientists and selected based on priorities, and information regarding the approved targets were used by mission operations in Japan. The basic procedures of the GO program were established.
    \item[1b-ASCA$^{+}$:] All data products were well managed and distributed to GOs. The backbone of the procedure for processing and archiving observation data was established.
    \item[2a-ASCA$^{+}$:] The core algorithms in the analysis software were well verified via on-ground calibration measurements before launch by instrument teams (ITs). Analysis tools using these algorithms and the calibration database were delivered to GOs. The concept was established in this mission.
    \item[3a-ASCA$^{+}$:] As a part of the GO program, user support activities were established. 
    \item[4a-ASCA$^{+}$:]  Collaboration between Japan and US was established on the {\it ASCA} science operations and was well organized especially on the development of the public software and the calibration database.
    \item[4b-ASCA$^{+}$:] The instrument team members in Japan performed ground calibrations while scientists both in Japan and US performed continuous in-orbit calibrations, which delivered good calibration accuracy.
    \end{description}
\end{quote}
However, the following items can be regarded as negative lessons provided by this past mission.
\begin{quote}
    \begin{description}
    \item[2b-ASCA$^{-}$:] The instrument team members developed their own tools for analyzing the ground calibration data, which sometimes provide better results than the public analysis software released as part of item {\bf 2a-ASCA$^{+}$} in the early GO phase. 
    Since these tools were not initially made public, they became referred to as "animal software" because of the unfairness of the analyses from the viewpoint of a public observatory, although this unfair situation was resolved in the final version of the products.
    \item[3b-ASCA$^{-}$:] GO support was provided only in the US by the US {\it ASCA} Guest Observer Facility (GOF), but not in Japan, although instrument teams in Japan provided deep support for the GOF activity. The interface to GOs existed only in the US.
    \end{description}
\end{quote}

\subsection{Lessons Learned from {\it Suzaku}}
\label{sec:ll:suzaku}
The {\it Suzaku} mission was the fifth in the series of the Japanese X-ray satellites in collaboration between JAXA and NASA \cite{2007PASJ...59S...1M}, and was launched in 2005 carrying the High-throughput X-ray Telescope, X-ray Imaging Spectrometer CCD cameras, and non-imaging Hard X-ray Detector.
The science operations members of {\it Suzaku} tried to utilize the positive lessons from {\it ASCA} (i.e., on the GO program and data distribution {\bf 1a-ASCA$^{+}$} and {\bf 1b-ASCA$^{+}$}, the software development {\bf 2a-ASCA$^{+}$}, the GO support {\bf 3a-ASCA$^{+}$}, and the PVO activities {\bf 4a-ASCA$^{+}$} and {\bf 4b-ASCA$^{+}$}) and fix the negative situations ({\bf 2b-ASCA$^{-}$} and {\bf 3b-ASCA$^{-}$}). They successfully fixed these using {\bf 2ab-Suzaku$^{+}$} and {\bf 3b-Suzaku$^{+}$}, respectively, which are summarized below.
\begin{quote}
    \begin{description}
    \item[2ab-Suzaku$^{+}$:] In order to keep the positive situation {\bf 2a-ASCA$^{+}$} and solve negative situation {\bf 2b-ASCA$^{-}$} (i.e., avoiding animal software), the public tools and tools for ground calibration measurements were designed to have the core algorithms for calculating variables such as time, pulse-height invariant (PI\cite{2005ITNS...52..902T}), and grade, shared as software libraries.
    The instrument team members developed and verified these core libraries via hardware development, and the same libraries were smoothly exported into public tools that could be used by the GOs.
    Therefore, the public tools were well verified and well calibrated.
    \item[3b-Suzaku$^{+}$:] In order to improve situation {\bf 3b-ASCA$^{-}$}, a {\it Suzaku} Help Desk in Japan (RIKEN)\cite{2007PThPS.169..312T} were operated in addition to the {\it Suzaku} GOF in the US. 
    Since several members of the {\it Suzaku} Help Desk also belong to the instrument teams in Japan, these two bodies had a strong potential for solving questions from GOs very quickly because of their tight connection to the operation team and developers of instruments.
    \end{description}
\end{quote}
In addition, starting from {\it Suzaku}, communication was established with other X-ray missions in terms of calibration activities (keeping {\bf 4b-ASCA$^{+}$}), as indicated as {\bf 4c-Suzaku$^{+}$} below.
\begin{quote}
    \begin{description}
    \item[4c-Suzaku$^{+}$:] The {\it Suzaku} instrument members participated from the beginning in cross-calibration activities in the International Astrophysical Consortium for High Energy Calibration (IACHEC) \cite{2020SPIE_IACHEC}.
    \end{description}
\end{quote}
However, the following two negative points related to {\bf 2ab-Suzaku$^{+}$} and {\bf 3b-Suzaku$^{+}$} arose in the {\it Suzaku} Science Operations, and were left as open issues for the next mission ({\it Hitomi}).
\begin{quote}
    \begin{description}
    \item[2c-Suzaku$^{-}$:] Software development by the instrument teams in {\bf 2ab-Suzaku$^{+}$} caused a) unexpected software freezes and b) delays in the delivery schedule, because a) not all the instrument members were experts on programming, and b) the first priority of the instrument teams was the delivery and maintenance of the detector itself, with software development having a lower priority.
    \item[3c-Suzaku$^{-}$:] The members of the {\it Suzaku} Help Desk in Japan ({\bf 3b-Suzaku$^{+}$}) were not appointed by the agency and had no special data-access permission. Therefore, the tasks were performed on a best-effort basis and sometimes activity stopped because of other business. 
    \end{description}
\end{quote}

\subsection{Lessons Learned from {\it Hitomi}}
\label{sec:ll:hitomi}
The {\it Hitomi} mission was the sixth in the series of Japanese X-ray satellites developed at JAXA in collaboration with NASA and Japanese and Canadian institutions with contribution from the ESA \cite{2018JATIS...4b1402T}, and carried an micro X-ray calorimeter array and X-ray CCD cameras on the focal plane of X-ray mirrors, as well as hard X-ray instruments with hard X-ray mirrors and a soft gamma-ray detector.
The satellite was successfully launched in February 2016, but contact with the spacecraft was lost in March 2016 owing to problems in the bus system before the performance verification (PV) phase.
Therefore, most of the science operations after launch were not activated, as indicated by {\bf 1a-Hitomi$^\pm$}, {\bf 3-Hitomi$^\pm$}, and {\bf 4-Hitomi$^\pm$} below.
\begin{quote}
    \begin{description}
    \item[1a-Hitomi$^\pm$:] Opportunity for calling for GO proposals was canceled, although the distribution of the in-orbit data was completed (keeping {\bf 1b-ASCA$^{+}$}).
    \item[3-Hitomi$^\pm$:] GO support helpdesks were prepared but not activated.
   \item[4b-Hitomi$^\pm$:] Calibration of instruments was performed on the ground and partially in orbit during the commissioning phase, and the results were released in the calibration database. 
    \end{description}
\end{quote}
In the science operations of {\it Hitomi}, positive lessons from {\it ASCA} and {\it Suzaku} were kept in the activities under the above situations, including participation in the IACHEC ({\bf 4c-Suzaku$^{+}$}).
In parallel, the science operations members tried to solve negative item {\bf 2c-Suzaku$^{-}$} (leaving {\bf 3c-Suzaku$^{-}$} open when the mission terminated), and it was solved as {\bf 2c-Hitomi$^{+}$} with {\bf 2d-Hitomi$^{-}$} left as an open issue as shown below.
\begin{quote}
    \begin{description}
    \item[2c-Hitomi$^{+}$:] In order to avoid {\bf 2c-Suzaku$^{-}$} (unexpected software freeze and schedule delay), a specific team was defined for the development of software and the calibration database. The team was the {\it Hitomi} software and calibration team (SCT) and was independent from the instrument teams and consisted of scientists and programmers. 
    As a result, there were no delays in the schedule of software preparation and no delay in the release of tools and the calibration database. 
    The products were well calibrated using instrument-specific methods\cite{2018JATIS...4a1206T, 2018JATIS...4b1407L, 2018JATIS...4b1406E, 2018PASJ...70...20T, 2018PASJ...70...18K, 2016RScI...87kD503E, 2018JATIS...4a1213I, 2018PASJ...70...19M, 2018JATIS...4a1210M, 2018JATIS...4b1409H}, because all the algorithms were imported into the analysis software and the latest calibration information was quickly released in the database\cite{2018JATIS...4a1207A}.
    \item[2d-Hitomi$^{-}$:] In order to achieve {\bf 2c-Hitomi$^{+}$}, there were many interactions among the software and calibration team, instrument teams, and operation teams, which were spread across multiple agencies. Therefore, many more tasks than expected were required in order to manage tasks for science operations such as the schedule, manpower of activities, and interfaces.  
    \end{description}
\end{quote}

\subsection{Recommendations for {\it XRISM} Science Operations}
\label{sec:ll:xrism}
In summary, from the lessons learned from {\it ASCA}, {\it Suzaku}, and {\it Hitomi} missions, two items labeled {\bf 2d-Hitomi$^{-}$} (team management issues) and {\bf 3c-Suzaku$^{-}$} (data access rights in users support) remain as open items for the {\it XRISM} Science Operations, with the remaining items recommended to remain unchanged. 
The two items are related to the management of manpower and the preparation of science operations before launch.

\section{Concept for the {\it XRISM} Science Operations}
\label{sec:concept}

Based on the lessons learned from the past X-ray missions and recommendations for {\it XRISM} operations (Section \ref{sec:ll}), we established the {\it XRISM} Operations Concept, as described in this section.

\subsection{Key points of the {\it XRISM} Operations Concept}
\label{sec:concept:summary}
Considering the recommendations from lessons learned from past X-ray missions (Section \ref{sec:ll}), the key points of the {\it XRISM} Operations Concept can be summarized into the following three items.
\begin{quote}
    \begin{enumerate}
        \item[{\bf OC01}] Clear division between spacecraft operations (hereafter, "mission operations") and science operations is required so that the scientists can concentrate on science operations.
        \item[{\bf OC02}] The plans for the operations (both mission and science operations), including the team structure and interfaces, should be defined in an early phase before launch. Similarly, training and actual operations should start before launch.
        \item[{\bf OC03}] All members of the operations (both mission and science operations) should be appointed by the agencies, and all activities, except for PVO activities (see Section \ref{sec:ll:summary}), should be performed as well-defined tasks with clear due dates, that continue to work until the end of the mission.
    \end{enumerate}
\end{quote}

\subsection{Task Division between Mission Operations and Science Operations}
\label{sec:concept:mot_sot}
In operations concept {\bf OC01}, operations tasks that require scientific decisions from the viewpoints of scientists are all assigned as {\it XRISM} Science Operations, and all other tasks are assigned as {\it XRISM} Mission Operations.
For example, the weekly negotiation of contact passes of ground stations, generation of daily operation commands, execution of the spacecraft simulator, down-link-and-command operations at ground stations, and checking the house-keeping telemetries from the spacecraft in real-time and off-line are assigned as tasks for mission operations. 
On the other hand, the handling of proposals from GOs, scientific scheduling of astrophysical objects, quick-look checks of science telemetries, instrument calibration activities\footnote{However, the calibration activities were shared among {\it XRISM}-internal groups, as described later in Section \ref{sec:plan}.}, and the management of daily data process and archive operations and user support activities are considered tasks for science operations.

We defined individual teams for performing  {\it XRISM} Mission Operations and {\it XRISM} Science Operations separately, which are called the Mission Operations Team (MOT) and Science Operations Team (SOT), respectively. 
Following the lessons of {\bf 2c-Hitomi$^{+}$}, these teams also need to be independent from the instrument teams. 
In addition, we defined the Science Management Office (SMO) to manage the overall {\it XRISM} Science Operations for deciding items regarding science operations. 
For example, the SMO handles activities such as calling for, reviewing, and selecting GO proposals, and approving targets for director time-of-opportunity (ToO) observations.

\subsection{Operation Phases and Team Structure}
\label{sec:concept:phase_team}
In operations concept {\bf OC02}, the operation phases of the {\it XRISM} are defined as follows.
\begin{quote}
    \begin{enumerate}
        \item Before Proto-Flight Test (PFT) Phase
        \item PFT Phase
        \item Launch Preparation Phase \& Launch Phase
        \item Initial Phase, which consists of the Critical Operations Period and Commissioning Period
        \item Nominal Operations Phase, which consists of the Initial-Calibration and Performance-Verification (PV) Period and Nominal Observation Period
        \item Latter Phase, which consists of the Latter Observation Period and Completion of Operations
    \end{enumerate}
\end{quote}
Based on concept {\bf C02}, the phase in which science operations starts should be the 2) PFT Phase, not the 5) Nominal Operations Phase after launch.
In other words, preparation of the mission and science operations should be completed before the PFT Phase on the ground, and the MOT and SOT shall start from the PFT Phase.
Therefore, we define the Mission Operations Planning Team (MOPT) for preparing the mission and science operations (e.g., construction of the detailed operations plan, preparation of the OTs and ground system), which is active from the 1) Before PFT Phase. 

According to operations concept {\bf OC03}, all members of the MOT and SOT should be appointed by the agencies (JAXA or NASA), and work on well-defined tasks under a managed schedule until the end of the mission, although members of the MOPT may be non-appointed members from universities as developers of tools in the Before PFT Phase. 
The SOT members consist of not only leader(s) and senior scientists, but also young scientists, referred to as Duty Scientists, who perform the actual {\it XRISM} Science Operations and are appointed by the agencies.
In our concept, the tasks of the Duty Scientists should be defined such as to provide a good career path for young scientists.
Note that the concept of the Duty Scientists is applied only on the science operations center in Japan as described in the next section \ref{sec:team}.

\section{Design of the Team and Management Structure of the {\it XRISM} Science Operations}
\label{sec:team}

Based on operations concepts {\bf OC01}, {\bf OC02}, and {\bf OC03} in Section \ref{sec:concept}, we defined the details of the structure of the SOT and the interfaces and task divisions among the subgroups of the {\it XRISM} team, which are described in this section.

\subsection{Interface Structure Between Subgroups and the SOT}
\label{sec:team:interface}
\begin{figure}[ht]
    \centering
    \includegraphics[width=0.85 \textwidth]{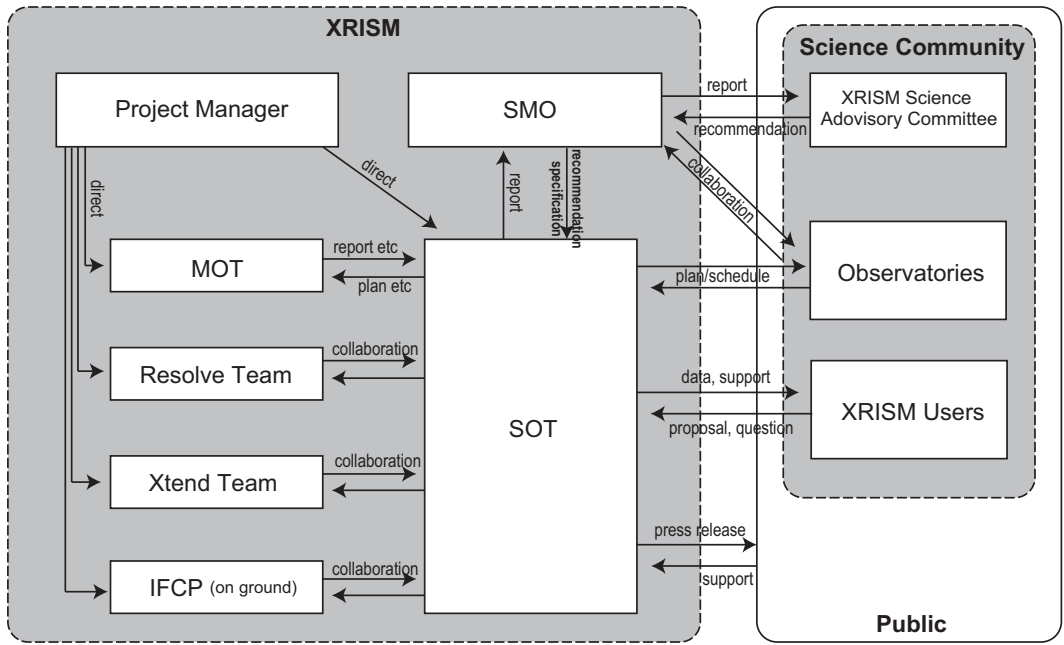}
    \caption{Interactions between the SOT and other internal subgroups in the {\it XRISM} team after the PFT Phase. The scientific community and public are also shown on the right.}
    \label{fig:team_interface}
\end{figure}

In addition to the SOT, MOT, and SMO described in Section \ref{sec:concept}, the {\it XRISM} team consists of the instrument teams, namely, the Resolve and Xtend teams, and the In-flight Calibration Planning Team (IFCP)\cite{2020SPIE_XRISM_IFCP}, which provides the detailed plans for in-orbit calibration observations before launch.
Interactions between these subgroups after the PFT Phase (Section \ref{sec:concept:phase_team}) are summarized in Figure \ref{fig:team_interface}. 

The SOT is directed by the Project Manager (PM), and works with the SMO, which provides recommendations and specifications for science operations as established in the concept study in Section \ref{sec:concept}.
The SOT does not communicate directly with the Science Advisory Committee, but rather through the SMO. 
The SOT communicates with the MOT regarding tasks such as the planning of spacecraft operation, verification of telemetry data, and reports. 
Several tasks such as in-orbit calibration observations and instrument performance monitoring are performed in collaboration with the Resolve, Xtend, and IFCP teams. 
The SOT communicates with GOs directly regarding the acceptance of proposals and user support activities.
Communications with other observatories (i.e., X-ray missions and/or observatories in other wavelengths), negotiations for in-orbit calibration campaigns, and/or multi-wavelength scientific programs are handled by the SMO, and communications on actual observation plans/schedules are handled by the SOT.
Similarly, decisions regarding press releases of scientific outputs or mission status are made by the SMO, and the actual work of the publications is done by the SOT members under the direction of the SMO.

The calibration activities of payload instruments consist of many steps, and the task divisions among the SOT, instrument teams, and IFCP team are defined as Table \ref{tab:task_division_calibration}.

\begin{table}[ht]
    \centering
    \caption{Task division among the SOT, ITs, and IFCP team on the calibration activities.}
    \begin{tabular}{ll}
        \hline 
        \multicolumn{1}{c}{\bf Calibration Tasks} &  \multicolumn{1}{c}{\bf Subgroup Name}  \\
        \hline 
        Preparation of calibration requirement      & ITs\\
        Review calibration requirement from viewpoint & SOT \\
        of flow down from mission science objectives& \\
        \hline 
        Ground calibration tests                 & ITs and \\
                                                 & bus company on timing system \\
        Analyses of ground calibration data      & ITs with SOT and \\
                                                 & bus company with SOT on timing system \\
        Prepare pre-launch calibration database  & ITs with SOT \\
        Release pre-launch calibration database  & SOT \\
        Preparation of in-orbit calibration plan & IFCP$^\dagger$ \\
        \hline 
        Schedule of in-orbit calibration targets & SOT \\
        Spacecraft operation of in-orbit calibration targets & MOT \\
        Analyses of in-orbit calibration data    & SOT and ITs \\
                                                 & with PVO contributor$^\ddagger$ \\
        Preparation of calibration database      & SOT and ITs \\
        Release calibration database             & SOT \\
        \hline 
    \end{tabular}
    
    {\scriptsize 
    $\dagger$ The IFCP team is active before launch.
    $\ddagger$ The PVO contributors are defined in Figure \ref{fig:team_management} in Section \ref{sec:team:management}.
    }
    \label{tab:task_division_calibration}
\end{table}

\subsection{SOT Structure and Task Divisions}
\label{sec:team:sot}
\begin{figure}[ht]
    \centering
    \includegraphics[width=0.85 \textwidth]{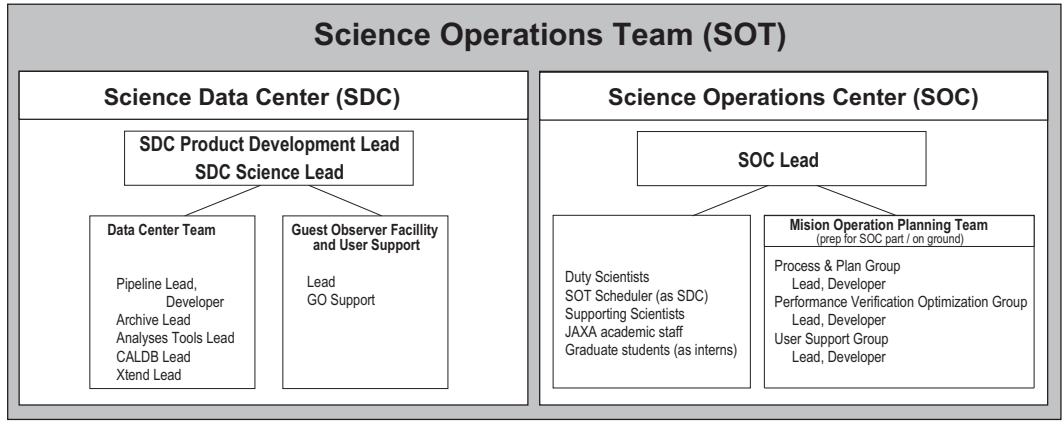}
    \caption{Structure of the SOT.}
    \label{fig:team_sot}
\end{figure}

The {\it XRISM} Science Operations are covered by JAXA, NASA, and the ESA.
Since tight collaboration between JAXA and NASA is required for preparation and maintenance of the data distribution ({\bf SO1} in Section \ref{sec:ll:summary}) and analysis software and the calibration database ({\bf SO2} in Section \ref{sec:ll:summary}), the SOT is designed to operate at the Science Operations Center (SOC) at JAXA and the Science Data Center (SDC)\cite{2020SPIE_XRISM_SDC} at NASA, as shown in Figure \ref{fig:team_sot}. 
Each center is made up of leads, scientists, and software engineers. 
In the SDC, the Product Development Lead and the Science Lead direct the SDC internal groups, namely, the Data Center Team, Guest Observer Facility, and User Support Group. 
The SOC is activated after the PFT Test Phase (see Section \ref{sec:concept:phase_team}), and the SOC Lead directs SOC members, such as the Duty Scientists defined in Section \ref{sec:concept:phase_team} and Supporting Scientists from the MOPT, after the PFT Phase.
Before launch, the science part of the MOPT is also under the direction of the SOC Lead, and consists of three groups: the Process and Plan Group, the PVO Group, and the User Support Group. 
In addition, the ESA operates the European Space Astronomy Center (ESAC) from the Before-PFT Phase and communicates with the SOC and SDC for the science operations in Europe. 

The task divisions among the three centers in JAXA, NASA, and ESA are defined in Table \ref{tab:task_division_sot},
summarized into four categories ({\bf SO1, SO2, SO3}, and {\bf SO4} in Section \ref{sec:ll:summary}).
The details of these tasks are described in the next Section \ref{sec:plan}.
\begin{table}[ht]
    \centering
    \caption{Task Division between SOC, SDC, and ESAC in the {\it XRISM} Science Operations.}
    \begin{tabular}{clll}
        \hline 
        {\bf Category} & \multicolumn{1}{c}{\bf SOC}
                 & \multicolumn{1}{c}{\bf SDC}
                 & \multicolumn{1}{c}{\bf ESAC} \\
        \hline 
        SO1 & Proposal support (JAXA)    & Proposal support (US/Canada)        &  Proposal support (Europe) \\
        (GO)& - Web pages for Submission & - Web pages for submission          & - Web pages for submission\\
            &                            & - Proposing systems and tools       & \\
            & - Proposal support         & - Proposal support                  & - Proposal support \\
            & - Supporting review process & - Supporting review process        & - Supporting review process \\ \cline{2-4}
        (Data)& Observation scheduling    & Observation scheduling              & \\ 
            & - Planning operations      & - Planning operations                & \\
            & - ToO handling            & - Planning tools and personnel       & \\ 
            & - MOT interface           &                                     & \\ \cline{2-4}
            &  Pre-pipeline process     & Pipeline process                    & (No data process)\\
            & - Format conversion       & - Filling calibrated columns           & \\
            & - Observation database    & - Metadata for archive             & \\
            &  Archiving at JAXA        & Archiving at NASA                   & (no archive) \\ 
            &  - Data archive           & - Data archive                      &  \\
            &  - Quick-viewing tool     & - Calibration database release      & - Quick-viewing tool\\ 
        \hline 
        SO2 &  Telemetry check& Development of analysis tools & \\
        (Software)& - Health check          & & \\
            & - Performance check       & Maintenance of calibration database& \\ 
        \hline 
        SO3 & User support (JAXA)       & User support (US/Canada)            & User support (Europe)\\
        (Support) 
            & - Handling GO questions   & - Handling GO questions             & - Handling GO questions \\
            & - User guide documents   & - User guide documents             & - Documentation\\ 
            & - Researcher webpages    & - HEASARC webpage                       & - ESAC webpage \\
            & - EPO support             & - EPO support                       & - EPO support \\
        \hline 
        SO4  & Calibration operations   & Calibration support                 & \\
        (PVO)& - Analyses               & - Analyses                          & \\
             & - Monthly plan           &                                     & \\
             \cline{2-4}
             & Performance check        & Performance check                   & Performance check\\
             & - Analysis threads       & - Analysis threads                  & - Analysis threads\\
             & - Daily/monthly check    & - Post-process development          & \\
             \cline{2-4}
             & Xtend Transient search   & & \\
        \hline 
    \end{tabular}
    \label{tab:task_division_sot}
\end{table}

\subsection{Management structure}
\label{sec:team:management}
\begin{figure}[ht] 
    \centering
    \includegraphics[width=0.75 \textwidth]{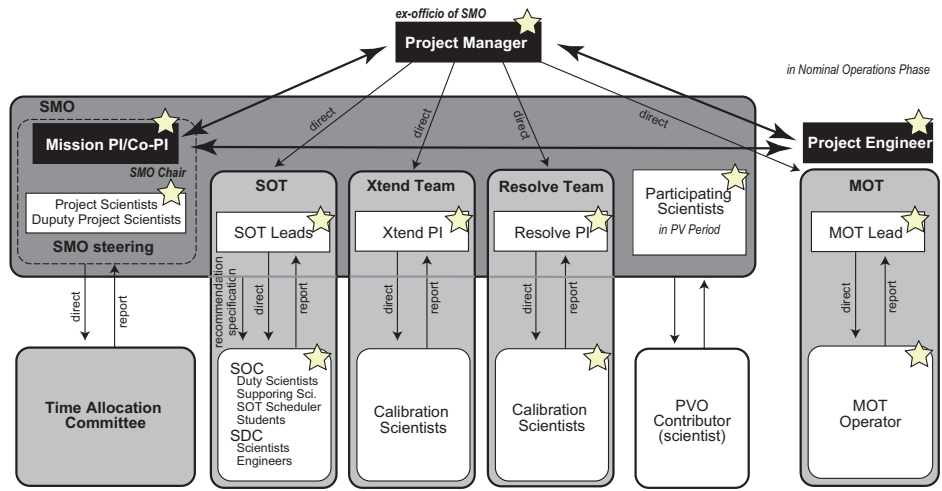}
    \caption{Organization structure of the {\it XRISM} Science Operations during the Nominal Operation Phase.
    }
    \label{fig:team_management}
\end{figure}

All specifications of the overall science operations are handled by the SMO, and therefore, the Mission Principal Investigator (PI) and co-PI (JAXA/NASA), Project Scientists (JAXA/NASA/ESA), Deputy Project Scientists (JAXA/NASA/ESA), and all the leaders of the internal subgroups (SOT and Resolve and Xtend teams) except for the MOT and IFCP team in Figure \ref{fig:team_interface} belong to the SMO Committee in the Nominal Operations Phase, as shown in Figure \ref{fig:team_management}.
The Project Managers both in JAXA and NASA also belong to the SMO as ex officio.
Before launch, leaders of the IFCP team, chairs of the science categories, which are active in the selection of the PV targets, and chairs of subgroups for specific topics such as atomic physics also participate in the SMO.

Following concept {\bf C03} in Section \ref{sec:concept:summary}, the members marked by $\largestar$ in Figure \ref{fig:team_management} are appointed by the agencies, namely, JAXA or NASA (or the ESA). 
Within the SOT, the SOC plans to have eight Duty Scientists (see Section \ref{sec:concept:phase_team}), as well as Support Scientists from the MOPT (Section \ref{sec:concept:phase_team}) and/or JAXA academic positions to help the Duty Scientists, as described in Section \ref{sec:team:sot}.
In the SDC, more than seven staff scientists and more than four software engineers will perform the science operations at NASA. 
All members of the SOC and SDC are appointed by JAXA and NASA, respectively, in the Nominal Operations Phase.

\subsection{Data Access Policy for Science Operations}
\label{sec:team:data_access}

In the science operations performed by the SOT, SOT members have access to all of the telemetry items including scientific properties in order to check the performance of the instruments and to make quick-look reports to GOs. These activities are limited to monitoring or checking of the instrument health and performance, and do not extend to performing scientific analyses of the scientific interests of scientists. 
The SOT members also check all the proposals approved by the SMO and their scientific justifications. 
While SOT members can access all the data and products to the extent that such access is required to perform their duties, they are required to maintain confidentiality of all scientific knowledge obtained in this context. 
The SOT members shall understand this data access policy in all of the science operations.


\section{Details of the Science Operations Plan}
\label{sec:plan}
Following operations concept {\bf OC02} (Section \ref{sec:concept:summary}), the detailed plan of the {\it XRISM} Science Operations is constructed as described in the following sections under the team structure defined in Section \ref{sec:team}, well before the launch during the Before PFT Phase (Section \ref{sec:concept:phase_team}).

\subsection{Summary of the Science Operations Scenario}
\label{sec:plan:summary}
All the tasks for the {\it XRISM} Science Operations are defined in terms of the four types of science operations defined in Section \ref{sec:ll:summary} (i.e., {\bf SO1, SO2, SO3}, and {\bf SO4}) 
in Table \ref{tab:tasks_by_steps}, which can be categorized into the following three-step operation flow.
\begin{quote}
    \begin{enumerate}
        \item[{\bf Step-1}:] Proposal and Planning Step (before observation)
        \item[{\bf Step-2}:] Telemetry Check and Data Processing and Archive Step (after daily spacecraft observation) 
        \item[{\bf Step-3}:] User Support and PVO Step (after distribution of the observational data to GOs)
    \end{enumerate}
\end{quote}
These steps are performed in parallel with observations, and are operated both by the SOT and MOT, with various timescales (once per year, monthly, weekly, daily, and continuous), as also shown in Table \ref{tab:tasks_by_steps}.
The tasks for the SOC in Japan (see Table \ref{tab:task_division_sot}) are shared among the Duty Scientists and the SOT Scheduler (who works on planning as a contribution from SDC staying at SOC; Figures \ref{fig:team_sot} and \ref{fig:team_management}) evenly by week or month. 
For example, one Duty Scientist performs the first task, which is then performed by another Duty Scientist the following week. 

\begin{table}[ht]
    \centering
    \caption{Tasks for {\it XRISM} Science Operations for the steps of observations (see the text in Section \ref{sec:plan:summary}) in the Nominal Operations Phase. Types of science operations are defined in Section \ref{sec:ll:summary}.}
    \begin{tabular}{ccllc}
    \hline 
     {\bf Step} & \multicolumn{1}{c}{\bf Type} & \multicolumn{1}{c}{\bf Category (frequency)} & 
                 \multicolumn{1}{c}{\bf Tasks}  & \multicolumn{1}{c}{\bf Subgroup} \\
    \hline 
     Step-1    
          & SO1   & GO proposal support  & Call for proposals, receive proposals &SOT \\
          & SO3   & (year)               & Support GO proposals                  &SOT \\
          & SO1   &                      & Support review process                &SOT \\
          & SO1   &                      & Handle approved target                &SOT \\ \cline{3-5}
          & SO1   & Observation planning & In-orbit calibration planning         &SOT \\
          & SO1   &  (monthly, weekly, daily)& ToO handling                      &SOT \\
          & SO1   &                      & Observation scheduling (long/short term)&SOT \\
          & SO1   &                      & Coordination of observations          &SOT \\
          & SO1   &                      & MOT I/F for operation command generation & SOT \\
    \hline 
    NA    & N/A   & Spacecraft operation &                                    &MOT\\
          &       &  (daily)             &                                    &\\
    \hline 
     Step-2    
          & SO2   & Telemetry check      & Process quick-look data (QLDP)     &MOT\\
          & SO2   &  (daily)             & Health check                       &MOT,SOT\\
          & SO4   &                      & Instrument performance check       &SOT\\ \cline{3-5}
          & SO1   & Data process         & Format conversion (PPL)            &MOT, SOT\\
          & SO1   & (daily)              & Calculate calibrated columns (PL)  &SOT\\
          & SO1   &                      & Observation database/Archive Metadata &MOT,SOT\\
          & SO2   &                      & Product check                      &SOT\\
          & SO1   &                      & User notification                  &MOT\\ \cline{3-5}
          & SO1   & Archive              & Data publication                   &MOT\\
          &       &  (daily)             &                                    &\\
  \hline 
     Step-3    
          & SO3   & User support         & Researcher webpages                    &SOT\\
          & SO3   &  (daily)             & Handling GO questions              &SOT\\
          & SO3   &                      & Documentation                     &SOT\\
          & SO3   &                      & EPO support                        &SOT\\ \cline{3-5}
          & SO4   & PVO activities       & Calibration analyses               &SOT\\
          & SO4   &  (daily)             & Monthly performance check          &SOT\\
          & SO4   &                      & Development of Analysis threads and tools &SOT\\
          & SO4   &                      & Xtend transient search             &SOT\\
    \hline 
    \end{tabular}
    \label{tab:tasks_by_steps}
\end{table}

\subsection{Details of Tasks in Step-1 Before Observation}
\label{sec:plan:plan}
Most of the tasks in Step-1 before observation are of Type {\bf SO1} (defined in Section \ref{sec:ll:summary}), which can be divided into the following two categories.
The details are as follows.

\begin{quote}
    \begin{itemize}
        \item {\bf GO proposal support} 
        \begin{itemize}
            \item The SOT supports calls for proposals by the SMO and receives proposals from GOs.
            During proposal acceptance, the SOT supports GOs with preparation of proposals ({\bf SO3}).
            \item After the review process, the SOT gets a prioritized approved target list from the SMO. 
            In parallel, the in-orbit calibration objects are merged into the list.
            The SOT puts information regarding approved targets and calibration objects into the observation database and opens the list via the webpages of the researchers.
        \end{itemize}
        \item {\bf Observation planning}
        \begin{itemize}
            \item After the SOT obtains the list of targets, the SOT Scheduler (defined in section \ref{sec:plan:summary}) generates a long-term operation plan taking into account the spacecraft operational constraints. 
            \item Using the long-term plan as a guide, the SOT generates a more detailed short-term observation schedule weekly and prepares the detailed plans for observations for the week.
            \item During preparation of the observation details, the SOT notifies the observation PI of the plan, and negotiates the hardware configuration with the instrument teams. 
            \item In addition to the planned objects, the SOT handles ToO proposals from GOs. 
            If the SMO approves a ToO proposal, the short-term operation plan is quickly updated and used for the spacecraft operations.
            \item Before spacecraft operations, the SOT acts as an interface to the MOT from SMO and GOs on scientific topics for the generation of operation commands to the spacecraft.
        \end{itemize}
    \end{itemize}
\end{quote}

\subsection{Details of Tasks in Step-2 After Observation}
\label{sec:plan:proc}
The tasks in Step-2 after observation are a mixture of Types {\bf SO1}, {\bf SO2}, and {\bf SO4} (Section \ref{sec:ll:summary}), and can be divided into the following three categories. The details are as follows.

\begin{quote}
    \begin{itemize}
        \item {\bf Telemetry check} 
        \begin{itemize}
            \item Telemetry from the spacecraft needs to be checked quickly after spacecraft operation.
            In order to quickly perform these telemetry checks before the official data processing for GOs, which takes about one or two weeks in total, a Quick-look Data Process (QLDP) is defined to generate products quickly.
            The QLDP simplifies the timing calibration, orbit determination, and attitude determination processes from the official data processing.
            \item The MOT executes the QLDP and performs the quick health checks of instruments using the housekeeping telemetry semi-automatically. This function checks every value of the attribute in the engineering housekeeping telemetry to verify the proper and safe operation of the observatory at all times. If the MOT find an anomalous telemetry for the spacecraft safety from this limit checks, they will respond immediately as an emergency operation. In any cases, the MOT reports the results to the SOT daily after the spacecraft operation.
            \item In addition to the engineering health checks by the MOT, further checks of the performance of payload instruments from a scientific viewpoint are also required for the SOT.
            The SOT uses the products from QLDP of not only the housekeeping telemetry but also the science telemetry, performs the pipeline-equivalent process to calculate data such as the time, coordinate, and energy information, and then checks the instrument performance.
        \end{itemize}
        \item {\bf Data Process} 
        \begin{itemize}
            \item After spacecraft operation, telemetry is converted into the FITS format\cite{2001A&A...376..359H} by the pre-pipeline (PPL) process, and then higher-level calculations, such as filling time, coordinate, and energy information (PI\cite{2005ITNS...52..902T}), are performed by the pipeline (PL) process.
            (Note that the details of the PPL and PL are described later in Section \ref{sec:prep:tools:summary}.)
            The products of the PPL and PL are archived and distributed to GOs. 
            Execution of the PPL and PL is performed by the MOT supported by the SOC and the SDC, respectively.
            \item The products of the PPL and PL are checked by the SOT before the distribution to GOs.
        \end{itemize}
        \item {\bf Archive} 
        \begin{itemize}
            \item The products of the PPL and PL checked by the SOT are archived both in JAXA and NASA archive centers, the Data ARchive and Transmission System (DARTS) and the High Energy Astrophysics Science Archive Research Center (HEASARC), respectively.
            \item When the products are archived, the SOT notifies the readiness to GOs.
        \end{itemize}
    \end{itemize}
\end{quote}

\subsection{Details of Tasks in Step-3 After Data Distribution}
\label{sec:plan:support}
The tasks in Step-3 after data distribution are of Types {\bf SO3} and {\bf SO4} (defined in Section \ref{sec:ll:summary}), which can be divided into the following two categories. The details are as follows.

\begin{quote}
    \begin{itemize}
        \item {\bf User Support} 
        \begin{itemize}
            \item The SOT prepares and operates webpages for GOs to provide information on GO proposals, operation schedules and logs, analysis documents, etc.
            Such researcher website for {\it XRISM} are prepared at the three agencies, JAXA, NASA, and ESA, separately but main contents are synchronized.
            \item The SOT operates the agency Help Desks for handling questions from GOs.
            \item The SOT prepares documents related to the data analyses of {\it XRISM}, such as analysis walkthrough, analysis manuals, and descriptions of instruments.
            \item Education and Public Outreach (EPO) activities are performed by other institutes in JAXA or NASA, and the SOT supports such activities for {\it XRISM}.
        \end{itemize}
        \item {\bf PVO Activities}
        \begin{itemize}
            \item  As defined in the task division of the calibration activities in Table \ref{tab:task_division_calibration}, the SOT calibrates payload instruments regularly with the instrument teams using the in-orbit calibration targets or trend archive data (i.e., non-scientific data obtained for performance monitoring, such as data during earth occultation of normal operations).
            \item In addition to the daily performance checks in Step-2, the SOT also monitors the instrument performance monthly.
            \item The SOT enhances the instrument performance (such as improving the pointing accuracy and tuning of the good time interval) by checking short-/long-term trends and correlations between telemetry items and performance parameters.
            The output of such performance enhancement activities are implemented as an analysis thread or a new analysis tool, which is provided to GOs via the researcher website or the software archive.
            \item 
            During the daily data checks in Step-2 (Table \ref{tab:tasks_by_steps}), the SOT carries out further analyses to search for possible new transient objects within the field of view of Xtend using the quick-look data products and the final products.
            If the SOT finds a transient and the SMO considers it worth reporting, the SOT posts a quick report to the Astronomers Telegram (ATEL, \url{http://www.astronomerstelegram.org/}) from quick-look data products, and updates the detailed information using the final products if necessary. 
            This activity is performed during the Initial-Calibration and PV Period under the permission of the observation PI during the Nominal Observation Period.
        \end{itemize}
    \end{itemize}
\end{quote}

\section{Preparation for Science Operations on the Ground}
\label{sec:prep}

Following the science operations plan in Section \ref{sec:plan},
the MOPT (Section \ref{sec:concept:mot_sot}) prepares science operations well before the launch along the timeline of the science operations (Section \ref{sec:prep:timeline}): i.e., the MOPT prepares the OTs for science operations (Section \ref{sec:prep:tools}), science operations manual, and website (Section \ref{sec:prep:support}), and performs the PVO activities from before launch (Section \ref{sec:prep:pvo}).
This section describes the preparation status at the end of the 1) Before PFT Phase (see Section \ref{sec:concept:phase_team}) on the ground.

\subsection{Timeline of the Science Operations Preparation}
\label{sec:prep:timeline}

In each operation phases defined in Section \ref{sec:concept:phase_team}, the MOPT and SOT prepare and/or perform the science operations following the timeline shown in Table \ref{tab:time_line}, respectively.
Before the launch, on ground, the MOPT prepares the descriptions for science operations, such as the science operations plan, the manual for the science operations, and the detail design of the operation tools (OTs), etc, and develops and verifies the OTs in the before PFT phase, and then trains the SOT members in the PFT phase.
The list of targets to be observed during the initial-calibration and PV period in the Nominal operations phase (Section \ref{sec:concept:phase_team}) is released during this phase. 
The PV target list has been released on Febrary 15, 2021.
After the launch of the satellite, the SOT supports the critical and commissioning operations by the MOT and ITs during the initial operations phase, prepares the data process of the first light object at the end of commissioning period, and, after that, performs the nominal operations in the Normal Operations Phase. 

\begin{table}[htb]
   \centering
   \caption{Timeline of the {\it XRISM} science operations.The type of science operations are defined in Section \ref{sec:ll:summary}, the CP($\dagger$) represents the calibration plan of instruments, and the ERS($\ddagger$) represents the Early Release Science targets in the PV period.)}
   \begin{tabular}{llllll}
     \hline 
     \multicolumn{1}{c}{{\bf Phase}}
                 & \multicolumn{1}{c}{\bf XRISM}    
                 & \multicolumn{1}{c}{\bf SO1}    
                 & \multicolumn{1}{c}{\bf SO2}     
                 & \multicolumn{1}{c}{\bf SO3}      
                 & \multicolumn{1}{c}{\bf SO4}  \\
                 &\multicolumn{1}{c}{\bf }   
                 &\multicolumn{1}{c}{\bf (GO, Data)}   
                 &\multicolumn{1}{c}{\bf (Software)}    
                 &\multicolumn{1}{c}{\bf (Support)}      
                 &\multicolumn{1}{c}{\bf (PVO)}      \\
     \hline 
     {\bf Before PFT} 
                & Reviews      & \multicolumn{3}{l}{ Prepare science operations plan and manual} & Review CP$^\dagger$\\
                &              & \multicolumn{3}{l}{ Design and verify OTs} & \\
                & Release PV list &                &                & Open webpages      & Prepare IFCP  \\
    \hline 
     {\bf PFT} 
                & PFT          & Verify OTs     & Process data    & Update Guide & \\
                &              & Update manual  & Update manual  & Update webpages   & \\
                & end-to-end   & \multicolumn{4}{l}{ Training Science Operations members} \\
    \hline 
     {\bf Launch} 
                & launch       & \multicolumn{2}{l}{Preparation of call for proposal} & \multicolumn{2}{l}{Update webpages} \\
                &              & \multicolumn{4}{l}{Release 1st version of {\it XRISM} software and calibration database }\\
    \hline 
    {\bf Initial} & \multicolumn{5}{l}{\bf Critical Period} \\
                &            & Internal process & Check telemetry   & & \\ \cline{2-6}
                & \multicolumn{5}{l}{\bf Commissioning Period}  \\
                &             & Internal process & Check telemetry  & Update webpages & \\
                & First light & Process data     &                   & Press release & \\
    \hline 
     {\bf Nominal} & \multicolumn{5}{l}{\bf Initial-Calibration/PV Period} \\
                &             & Process data      &Check telemetry   &            & Calibration \\ 
                & ERS$^\ddagger$ & Release ERS &                  & Update webpages & Search transient\\ 
                &             & \multicolumn{4}{l}{Preparation of call for proposal} \\ \cline{2-6}
                & \multicolumn{5}{l}{\bf Nominal observation Period} \\
                &             & Process data     & Check telemetry  & Update webpages & Calibration\\
                &             & Release PV/GO &                  & Helpdesk   & Search transient\\ 
                &             & \multicolumn{4}{l}{Preparation of call for proposal} \\
                &             & \multicolumn{4}{l}{Release/update {\it XRISM} software and calibration database } \\
    \hline 
     {\bf Latter} & Observation & \multicolumn{4}{l}{Continue the activities in the Nominal Operations Phase} \\
    \hline 

    \end{tabular}
    \label{tab:time_line}
\end{table}

\subsection{Tools for Science Operations and Detailed Designs}
\label{sec:prep:tools}
The tools and database required for the {\it XRISM} Science Operations by the steps (Section \ref{sec:plan:summary})  are summarized in Table \ref{tab:prep_tools}.
The responsibilities for these tools/databases in the subgroups within the SOT are also shown.
Among these 10 OTs, the proposal submission tools, planning tool, PL, calibration database, and analysis tools ({\bf OT01}, {\bf OT02}, {\bf OT06}, {\bf OT07}, and {\bf OT09}, respectively) are developed by the SDC, and the details of these OTs are described in Loewenstein {et.\ al.\ }(2020)\cite{2020SPIE_XRISM_SDC}.
Hereafter, this paper describes the details of the observation database, QLDP, PPL, and archive quick-viewing tool ({\bf OT03, OT04, OT05, OT08}) in Section \ref{sec:prep:tools} and the conversion tools for the researcher webpages ({\bf OT10}) in Section \ref{sec:prep:support}.

\begin{table}[ht]
    \centering
    \caption{List of science OTs and databases. Steps are defined in Section \ref{sec:plan:summary}.}
    \begin{tabular}{clll}
    \hline 
    Tool ID
    & \multicolumn{1}{c}{\bf Tools/Database}             
    & \multicolumn{1}{c}{\bf Step}    
    & \multicolumn{1}{c}{\bf SOT subgroups} \\
    \hline 
    OT01 & Proposal submission tools  & Step-1  &  SDC, (SOC, ESAC as users) \\
    OT02 & Planning tools              & Step-1  &  SDC with SOC \\
    OT03 & Observation database       & Step-1,2&  SOC \\
    OT04 & QLDP    & Step-2  &  SOC \\
    OT05 & PPL & Step-2  &  SOC \\
    OT06 & PL      & Step-2  &  SDC \\
    OT07 & Calibration database       & Step-2,3&  SDC with SOC and ITs \\
    OT08 & Archive quick-viewing tool & Step-2  &  SOC, (HEASARC), ESAC \\
    OT09 & Analysis tools             & Step-3  &  SDC with SOC and ITs \\
    OT10 & Conversion tools for researcher webpages & Step-1, -2, and -3 &  SOC\\
    \hline 
    \end{tabular}
    \label{tab:prep_tools}
\end{table}

\subsubsection{Pre-pipeline and Pipeline Process}
\label{sec:prep:tools:summary}
\begin{figure}[ht]
    \centering
    \includegraphics[width=0.50 \textwidth]{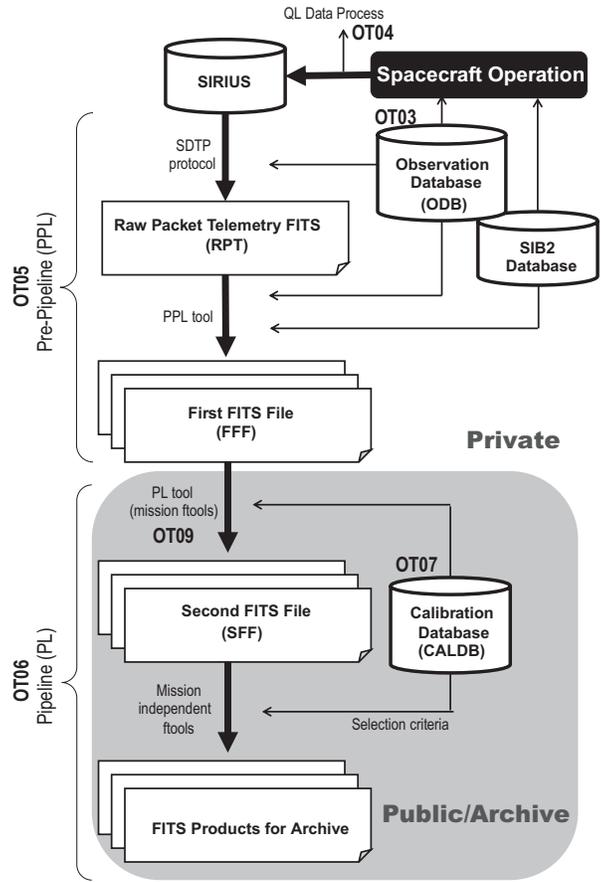}
    \caption{Schematic view of the flow of data processing for the {\it XRISM} (and {\it Hitomi, Suzaku}) in Step-2 and OTs.}
    \label{fig:flow_ppl_pl}
\end{figure}

Since the raw telemetry from the spacecraft is a collection of space packets, which are unreadable by the standard analyis tools used in high energy astronomy, they need to be converted into the standard FITS format\cite{2001A&A...376..359H} for distribution to GOs, as described in Angelini {et al.\ }(2018)\cite{2018JATIS...4a1207A}.
In addition, the GOs need calibrated information on variables such as time, coordinates, and pulse height invariant (PI\cite{2005ITNS...52..902T}; energy information), which are filled in by the data processing.
This corresponds to the the functions of a) format conversion and b) filling calibration columns.
The data processing is divided into two steps, PPL ({\bf OT05} in Table \ref{tab:prep_tools}) and PL ({\bf OT06} in Table \ref{tab:prep_tools}) as shown in Figure \ref{fig:flow_ppl_pl} to archive functions a) and b), respectively.

The raw telemetry from the spacecraft is stored in the SIRIUS database, and all the information regarding approved targets, instrument configuration, and other spacecraft information are stored in the observation database (ODB in Figure \ref{fig:flow_ppl_pl}; {\bf OT03} in Table \ref{tab:prep_tools}).
The PPL accesses the SIRIUS database via the Space Data Transfer Protocol (SDTP)  to retrieve the telemetry and first converts the telemetries into a raw packet telemetry (RPT) file, which is a simple dump of the series of space packets in the variable-length FITS format, using the information from the observation database ({\bf OT03}). 
In the second step, the PPL interprets the telemetry attributes in the space packets using the telemetry-description database, shown as the "Spacecraft Information Base version 2 (SIB2) database" in Figure \ref{fig:flow_ppl_pl}, and converts the RPT into the First FITS Files (FFFs) with meaningful columns.
The FITS header keywords of FFFs represent the instrument configurations as identified from the telemetry and observation database. 
After the PPL process, the PL fills in the calibration columns, such as time, coordinates, and PI, using the FITS tools called {\it ftools} in the HEASOFT {\it XRISM} package released from HEASARC by using the calibration database (denoted as "CALDB" in Figure \ref{fig:flow_ppl_pl}; {\bf OT07} in Table \ref{tab:prep_tools}), and stores them in the Second FITS Files (SFFs). 
The PL process then continues to extract the cleaned-event FITS for analyses from the SFFs by deleting low-quality events and by selecting good-time intervals. 

The key point of this procedure is that the FFFs have the same format as SFFs (i.e., FITS columns for time, coordinates, and PI are already prepared as blank columns in the FFF stage) and the CALDB and {\it ftools} are all distributed to GOs (i.e., public), so that the GOs can reprocess the SFF with the latest calibration information by themselves.
This concept was established in the {\it Suzaku} Science Operations, and also used in {\it Hitomi} successfully. The {\it XRISM} data process also follows this procedure.

The PPL requires inputs from the mission operation information, such as the definition of the telemetry format, the orbital estimation, the attitude determination, the time calibration, etc. 
Therefore, the PPL software for the {\it XRISM} mission is prepared and executed by the SOC at JAXA, where the mission operations are performed and the operation information is easily accessible from the SOT.
The FFFs are then sent via a data transfer system protocol (DTS\footnote{\url{https://heasarc.gsfc.nasa.gov/dts/}}) to the SDC, and processed in the PL at the SDC, as already described in the task division (Section \ref{sec:team:sot}).

\subsubsection{Design of Tools for the PPL and QLDP}
\label{sec:prep:tools:ppl}
\begin{figure}[ht] 
    \centering
    \includegraphics[width=0.85 \textwidth]{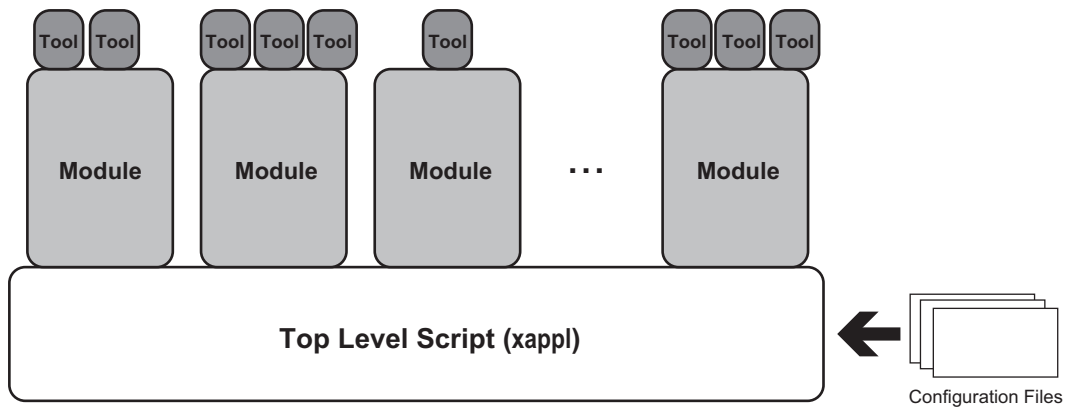}
    \caption{Structure of the PPL and QLDP.}
    \label{fig:ppl_structure}
\end{figure}

Since the QLDP ({\bf OT04} in Table \ref{tab:prep_tools}; Section \ref{sec:plan:proc}) is the simplified version of the PPL ({\bf OT05}), these tools can be shared with each other just by switching the execution mode.
The PPL and QLDP are designed to have a structure consisting of three stages of tool, modules, and top-level script, as shown in Figure \ref{fig:ppl_structure}, and the difference between PPL and QLDP is designed to be absorbed in the top-level script.
A tool is the smallest unit of the software code, and a module is a collection of tools to achieve one function (for example, generation of RPT, generation of time calibration fits, etc.).
The top-level script controls the process flow of multiple modules using the configuration files, with which the detail flow of PPL or QLDP are described.

Since {\it XRISM} is a recovery mission for {\it Hitomi}, the tools have already been developed and verified and can be reused for {\it XRISM}.
However, the {\it Hitomi} PPL is not easy to maintain because the {\it Hitomi} SCT (section \ref{sec:ll:hitomi}) were forced to use it during the Commissioning Phase for the unplanned spacecraft problems, and the {\it Hitomi} PPL has many patches for complex hardware modes during the Commissioning Phase, even though it was well designed for use in the Nominal Operations Phase.
Therefore, the MOPT decided to re-organize the PPL flow diagram and to newly develop the modules and top-level script for {\it XRISM}.

\begin{figure}[ht]
    \centering
    \includegraphics[width=0.6 \textwidth]{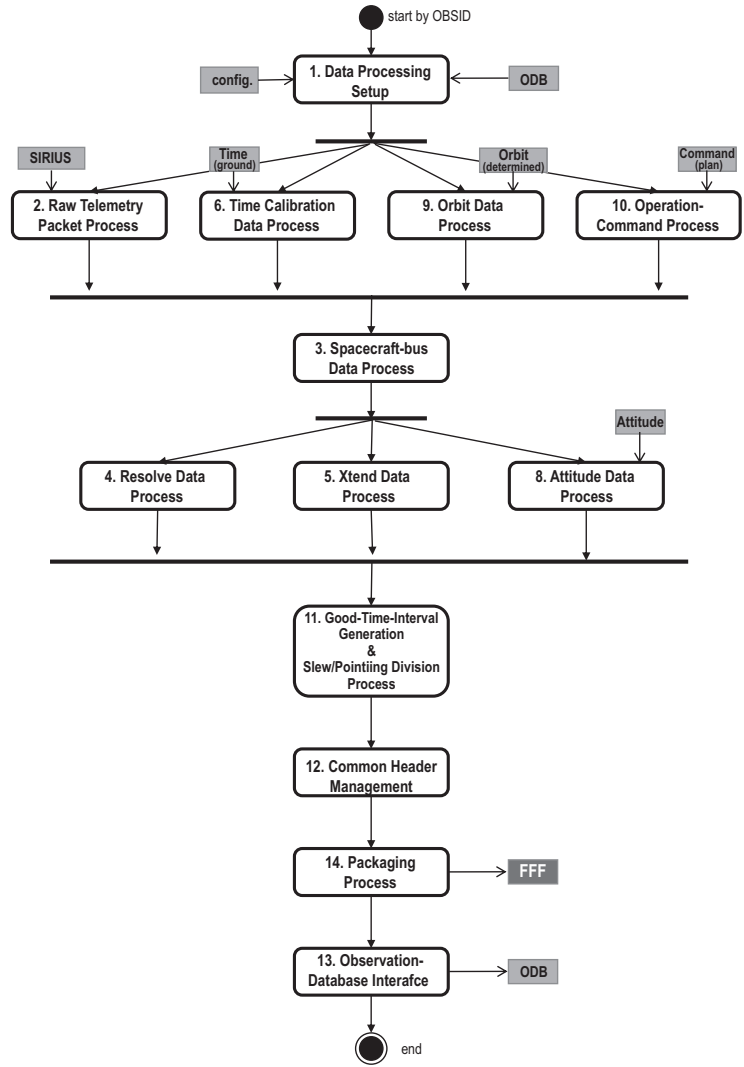}
    \caption{Flowchart of the PPL process.}
    \label{fig:ppl_flow_ppl}
\end{figure}

\begin{figure}[ht]
    \centering
    \includegraphics[width=0.6 \textwidth]{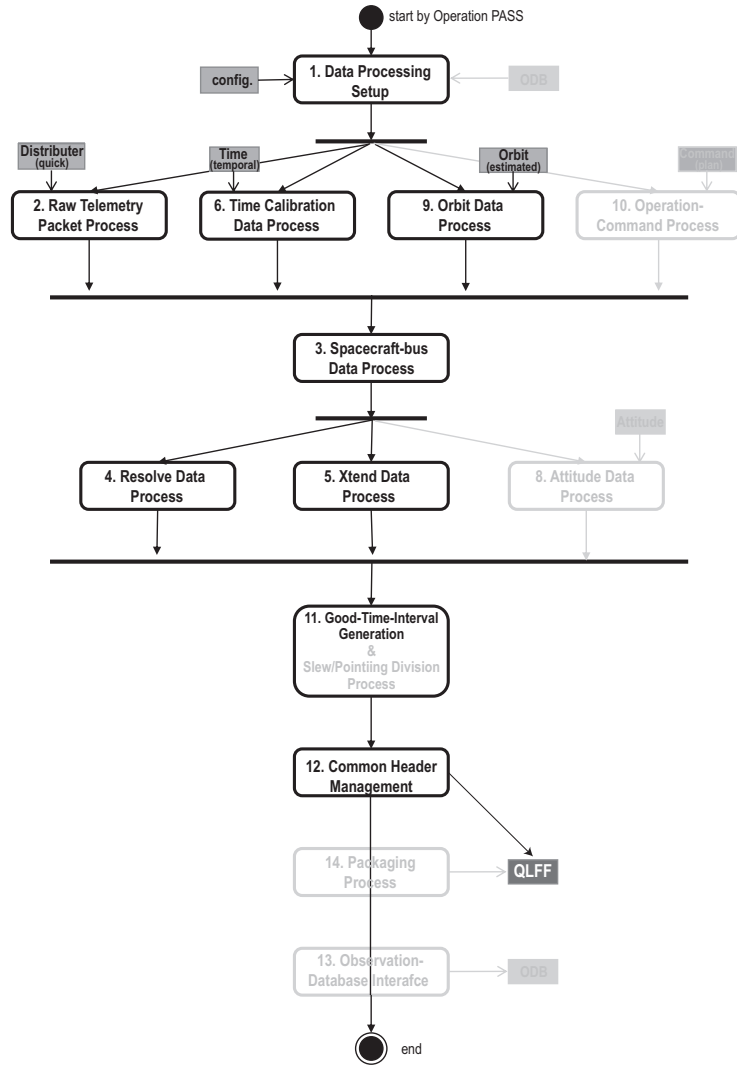}
    \caption{Same as Figure \ref{fig:ppl_flow_ppl}, but for QLDP. Gray colors represent the modules omitted in QLDP.}
    \label{fig:ppl_flow_ql}
\end{figure}

In detail, the MOPT defined the following 14 modules corresponding to the 14 functions required for the PPL and QLDP of {\it XRISM}.
Using these modules, the typical flow diagrams for the PPL and QLDP in the flight configuration are designed as shown in Figures \ref{fig:ppl_flow_ppl} and \ref{fig:ppl_flow_ql}, respectively.
In the QLDP, several modules are omitted in the process flow and the access point to retrieve the telemetry via SDTP is different from that for PPL, as well as the inputs for the time assignment tool and orbital-file generation tool.
In addition, the MOPT also identified 17 use cases for the ground tests and operations in orbit.
\begin{quote}
    \begin{enumerate}
        \item Data processing setup module
        \item Raw telemetry packet processing module
        \item Spacecraft-bus data processing module
        \item Resolve data processing module
        \item Xtend data processing module
        \item Time calibration data processing module
        \item Time assignment process/library
        \item Attitude data processing module
        \item Orbit data processing module
        \item Operation-command data processing module
        \item Good-time-interval generation and slew/pointing-division module
        \item Common header management module
        \item Observation-database interface module
        \item Package processing module
    \end{enumerate}
\end{quote}


\subsubsection{Design of Tools for the Archive Quick-Viewing Tool}
\label{sec:prep:tools:archive}
For the data archive for {\it XRISM}, it is important to define the division of tasks between the {\it XRISM} project and the archive centers of the agencies. 
In the archive activity at JAXA, the MOPT defined the task division between the SOT and the Center for Science Satellite Operation and Data Archive (C-SODA) at ISAS/JAXA, as summarized in Table \ref{tab:task_division_archive}.
In principle, content is provided by the {\it XRISM} project and preparation of mission-independent systems and infrastructure are made by C-SODA at ISAS/JAXA.

\begin{table}[ht]
    \centering
    \caption{List of tasks for archiving {\it XRISM} data at JAXA and division of tasks between SOT and C-SODA.}
    \begin{tabular}{llll}
        \hline 
        \multicolumn{1}{c}{\bf Category} & 
        \multicolumn{1}{c}{\bf Task} &
        \multicolumn{1}{c}{\bf Team}\\
        \hline 
        Data preparation &  Preparation of FITS files        & {\it XRISM} SOT \\
                         &  Definition of distribution range & {\it XRISM} SOT \\
                         &  Definition of proprietary period & {\it XRISM} SOT \\
                         &  Preparation of data storage      & JAXA C-SODA \\
                         &  Preparation of public webpages        & JAXA C-SODA \\
        \hline 
        Project introduction page & Preparation of contents  & {\it XRISM} SOT \\
                         &  Preparation of public webpages        & JAXA C-SODA \\
        \hline 
        Public data list &  Generation of the list and HTML  & {\it XRISM} SOT \\
                         &  Preparation of public webpages        & JAXA C-SODA \\
        \hline 
        Data search      &  Preparation of search engine  & JAXA C-SODA \\
                         &  Preparation of metadata         & {\it XRISM} SOT \\
                         &  Installation of {\it XRISM} metadata  & JAXA C-SODA \\
        \hline 
        Quick viewing    &  Preparation of quick-viewing system & JAXA C-SODA \\
                         &  Preparation of HiPS format of {\it XRISM} &  {\it XRISM} SOT \\
                         &  Installation of {\it XRISM} HiPS data  & JAXA C-SODA \\
        \hline 
    \end{tabular}
    \label{tab:task_division_archive}
\end{table}

Among the tasks by the SOT, the data preparation tasks (Table \ref{tab:task_division_archive}) are performed by the PPL and related tools {\bf OT05}, and the project introduction page and public data list tasks are performed manually by the SOT.
Therefore, additional OTs for the archive activity are a generation tool for a) metadata for data search and b) hierarchical progressive survey (HiPS) data for quick viewing, which are identified as {\bf OT08} in Table \ref{tab:prep_tools}.

\subsection{Preparation for User Support}
\label{sec:prep:support}
Researcher webpages are required for communication with GOs for the user-support activities listed in Table \ref{tab:tasks_by_steps}, and are planned to be operated in three centers: SOC, SDC, and ESAC. 
In the first version, the following content is listed on the researcher webpages at SOC.
\begin{quote}
    \begin{itemize}
        \item Top page, News and Announcements
        \item About XRISM (XRISM documents, workshops, publication list, resources)
        \item Proposer (GO proposal documents, response files, generic ToO request, approved target list)
        \item Observers (short-term/long-term operation plan, spacecraft operation log)
        \item Analysis (manual, link to archive web, link to download page for software/calibration database)
        \item {\it XRISM} Help Desk (FAQ, proposal plan support, analysis questions, {\it XRISM} workshops)
        \item Useful links (link to general public {\it XRISM} website, DARTS archive website, HEASARC website, ESAC website)
    \end{itemize}
\end{quote}

The MOPT prepares the web servers and the tools for filling the content of the pages of observation plan and spacecraft operation log semi-automatically. These tools are identified as {\bf OT10} in Table \ref{tab:prep_tools}. 
%
The JAXA researcher webpage was opened on 1 November 2020 at \url{https://xrism.isas.jaxa.jp/research/} and is used for announcements to GOs before launch and is to be activated on the science operations after launch.

\subsection{Preparation of PVO Activities}
\label{sec:prep:pvo}

In principle, all the PVO activities are performed by all the science members of the {\it XRISM} team before launch.
The items for the MOPT to prepare for the science operations in orbit are the detailed procedure for opening these efforts to GOs via the {\it XRISM} software, calibration database, and the analysis method, which are already covered in Section \ref{sec:plan}.

In order to perform the four science operations tasks described in Section \ref{sec:plan:support} and Table \ref{tab:tasks_by_steps}, the MOPT prepares the following items.

\begin{quote}
    \begin{itemize}
        \item {\bf Calibration analyses} \\
        As defined in Table \ref{tab:task_division_calibration}, the in-orbit calibration items and procedures are prepared by the instrument teams, and the SOT also analyzes the plan to observe the in-orbit calibration observations with the instrument teams. 
        Therefore, the MOPT prepares the training procedure for understanding the calibration items and procedures with the instrument team before launch.
        \item {\bf Monthly performance check} \\
        Daily and monthly checks of instrument performance require no special tools other than the standard analysis software. In this sense, the MOPT has no plan to prepare the tools before launch.
        If the MOPT identifies additional tools during the rehearsal of instrument operation on ground in the PFT Phase (Section \ref{sec:concept:phase_team}), the MOPT will prepare the tools from this phase.
        \item {\bf Development of analysis threads and tools} \\
        All of the standard analysis will be performed using the public standard tools ({\bf OT08} in Table \ref{tab:prep_tools}). 
        For the monthly or daily performance checks, the SOT tries to study and identify new proprieties or behaviors of instruments which affect the instrument performance. 
        If the SOT finds a way to enhance the instrument performance using these items, the SOT will implement the procedure as an analysis thread or prepare a new public tool using the newly found algorithm. 
        \item {\bf Xtend transient search} \\
        As described in Section \ref{sec:plan:support}, the SOT carries out further analyses to search for possible new transient objects within the field of view of the Xtend and posts a quick report to the ATEL, under the permission of the observation PIs. 
        The MOPT prepares the detailed procedure for this operation to obtain a quick response and the automatic search tool of transients. 
    \end{itemize}
\end{quote}

\section{Summary}
\label{sec:summary}

In preparation of science operations of the {\it XRISM} mission, we reviewed the lessons learned from past X-ray missions in Section \ref{sec:ll} to identify recommendations for the {\it XRISM} Science Operations (Section \ref{sec:ll:xrism}), which are considered as part of the operations concept (Section \ref{sec:concept}).
Based on the operations concepts, we designed the structure of the SOT, interfaces among subgroups, management structure, and data policy in Section \ref{sec:team} and established a detailed plan of the science operations as described in Section \ref{sec:plan}.
As the final step of preparation of science operations, we identified 10 OTs and developed them as summarized in Section \ref{sec:prep} before launch.

\acknowledgments
We thank Dr. Masa Sakano (WiseBabel Ltd.) for the detail design and implementation of the top-level-script of PPL, and Ms. Seiko Sakurai (Saitama University) for the implementation of the XRISM researcher’s website. 
We also thank Dr. Matteo Guainazzi, Dr. Jan-Uwe Ness (ESA), Dr. Katja Pottschmidt, and Dr. Tess Jaffe (NASA/GSFC) for the discussions on the user support activities in collaboration with ESAC and HEASARC. 
This work was supported in part by 
JSPS KAKENHI Grant Numbers JP18H04571 and JP20K04009 (YT),
JP19K14762 (MS), JP17K14289 (MN), JP20K20935 (SK and MT), 17K05392 (Y. T.), 
and NASA Grant Number 80NSSC20K0737.

\appendix
\section{Acronyms and Abbreviations}

Abbreviations and acronyms used in this paper are shown in Table \ref{tab:acronyms}.

\begin{table}[]
    \centering
    \caption{Acronyms and Abbreviations.}
    \begin{tabular}{llc}
    \hline 
     {\bf Abbreviations} & {\bf Acronyms} & {\bf Section} \\
     \hline 
ATEL & the Astronomers Telegram & section \ref{sec:plan:support} \\
C-SODA & the Center for Science Satellite Operation and Data Archive & section \ref{sec:prep:tools:archive}\\
CALDB & the calibration database & section \ref{sec:prep:tools:summary}\\
DARTS & the Data ARchive and Transmission System & section \ref{sec:plan:proc} \\
EPO   & Education and Public Outreach & section \ref{sec:plan:support} \\
ESA & the European Space Agency& section \ref{sec:intro} \\
ESAC & the European Space Astronomy Center & section \ref{sec:team:sot}\\
FFF & the First FITS Files& section \ref{sec:prep:tools:summary}\\
FITS & flexible image transport system& section \ref{sec:intro} \\
FOV & field of view& section \ref{sec:intro} \\
GO & Guest Observer & section \ref{sec:ll:summary}\\
GOF & Guest Observer Facility & section \ref{sec:ll:asca} \\
HEASARC & the High Energy Astrophysics Science Archive Research Center& section \ref{sec:plan:proc} \\
IACHEC & the International Astrophysical Consortium for High Energy Calibration & section \ref{sec:ll:suzaku} \\
ISAS & the Institute of Space and Astronautical Science & section \ref{sec:ll:asca} \\
IFCP & the In-flight Calibration Planning & section \ref{sec:team} \\
IT & instrument team & section \ref{sec:ll:asca} \\
JAXA & the Japan Aerospace Exploration Agency & section \ref{sec:intro} \\
MOPT & the Mission Operations Planning Team & section \ref{sec:concept:phase_team}\\
MOT & the Mission Operations Team & section \ref{sec:concept:mot_sot} \\
NASA & National Aeronautics and Space Administration& section \ref{sec:intro} \\
OC1 -- OC3 & {\it XRISM} Operations Concept & section \ref{sec:concept:summary} \\
OT & operations tool& section \ref{sec:prep:timeline} \\
PFT & Proto-Flight Test & section \ref{sec:concept:phase_team}\\
PI &  pulse-height invariant & section \ref{sec:ll:suzaku} \\
PL   & pipeline& section \ref{sec:plan:proc} \\
PM & the Project Manager& section \ref{sec:team:interface} \\
PPL  & the pre-pipeline & section \ref{sec:plan:proc} \\
PV & the performance verification & section \ref{sec:ll:hitomi} \\
PVO & Performance verification and optimization & section \ref{sec:ll:summary}\\
QLDP & Quick-look Data Process & section \ref{sec:plan:proc} \\
RPT & raw packet telemetry& section \ref{sec:prep:tools:summary}\\
SCT & {\it Hitomi} software and calibration team & section \ref{sec:ll:hitomi}\\
SDTP & the Space Data Transfer Protocol & section \ref{sec:prep:tools:summary}\\
SFF & the Second FITS Files& section \ref{sec:prep:tools:summary}\\
SIB2 & Spacecraft Information Base version 2& section \ref{sec:prep:tools:summary}\\
SMO & the Science Management Office & section \ref{sec:concept:mot_sot} \\
SO1 -- SO5 & Science Operations Type & section \ref{sec:ll:summary} \\
SOT & the Science Operations Team & section \ref{sec:concept:mot_sot} \\
ToO & time of opportunity & section \ref{sec:concept:mot_sot} \\
XRISM & The X-Ray Imaging and Spectroscopy Mission & section \ref{sec:intro} \\
     \hline 
    \end{tabular}
    \label{tab:acronyms}
\end{table}

%

\vspace{2ex}\noindent\textbf{ Yukikatsu Terada} is an associated professor at Saitama University, cross-appointed with Japan Aerospace Exploration Agency. He received his BS and MS degrees in physics, and a PhD degree in science from the University of Tokyo in 1997, 1999, and 2002, respectively. He is the leader of the Science Operations Team of the X-Ray Imaging and Spectroscopy Mission (XRISM) project now.

\vspace{1ex}
\noindent Biographies and photographs of the other authors are not available.


\end{spacing}
\end{document}